\documentclass{aastex631}

\shorttitle{Sunspots rotation}
\shortauthors{Tlatov and Tlatova}
\graphicspath{{./}{figures/}}

\begin{document}

\title{Differential Rotation of Individual Sunspots and Pores}

\author[0000-0002-6286-3544]{Andrei G. Tlatov}
\affiliation{Kislovodsk Mountain Astronomical Station of the Pulkovo observatory, Kislovodsk, Gagarina str. 100, 357700, Russia}


\author{Kseniia A. Tlatova}
\affiliation{Kislovodsk Mountain Astronomical Station of the Pulkovo observatory, Kislovodsk, Gagarina str. 100, 357700, Russia}

\begin{abstract}

The analysis of the rotation rate of individual sunspots and pores was performed according to the data of the processing of observations by the \textit{Solar Dynamics Observatory/Helioseismic and Magnetic Imager} (SDO/HMI)  in the period 2010\,--\,2024. Sunspots stood out in the images in the continuum. To accurately track the spots, we processed 5 images for each day. To determine the polarity of the magnetic field, we superimposed the contours of sunspots on observations of magnetic fields at the same time. This made it possible to track the movement of more than 210 thousand individual sunspots and pores. 
It is found that the rotation rate is influenced by the rotation  rate of the solar atmosphere and the systematic proper motions of the spots. Sunspots and pores of the leading polarity have a rate of meridional movement $\approx 2.4\%$ faster than spots of the trailing polarity. We also found that regular sunspots, which have umbra and penumbra, rotate $\approx 1.5\%$ faster than solar pores, in which penumbra is absent. The dependence of the rotation rate on these area is found. For sunspots with an area of $S> 10$ $\mu$hm, the rotation rate is practically independent of these area. Small sunspots, with an area of lower than $S< 10$ $\mu$hm, rotate $\approx 1.7\%$ slower.

\end{abstract}


\section{Introduction} \label{sec:intro}

The rotation rate of the Sun around its axis varies from many parameters. Different methods and different tracers used to track movement give different results \citep{Beck}.  Long-term measurements of the rotation rate of the upper layers of the solar atmosphere by averaging over longitude and time show that the rotation rate depends on latitude and depth \citep{Howe}.  Sunspots are often used to study long-term rotation variations.  Sunspots rotate at a different rate than non-magnetic plasma. The most available data for studying long-term variations in the rotation of the Sun are data on sunspot groups. Based on the analysis of Mount Wilson observations from 1921 to 1982 \citep{Howard}, the law of rotation of sunspot  groups  $\omega(\theta)=14.52-2.84  {\rm sin}^{2}\theta$  deg/day was found.  According to the processing of the Greenwich Photoheliographic Results 1874-1976 \citep{Balthasar1986}, the law of rotation was found $\omega(\theta)=14.557-2.85  {\rm sin}^{2}\theta$

In addition to latitude, the rotation rate of sunspots is influenced by other factors. A number of authors have found that the rotation rate decreases with the area of the sunspot groups \citep{Howard,Gupta, Nagovitsyn}. Rotation depends on the type of sunspot groups \citep{Balthasar1986}: C-groups exhibit high rate, and H- and J-spots are the smallest.  Some authors note that the  that the rotation of young sunspot groups occurs about 7\% faster than the rotation of older sunspot group \citep{Tuominen,  Balthasar,Pulkkinen}.  On the other hand, \citep{Javaraiah,Sivaraman} found that long-lived sunspot groups   accelerate these rotation over time. In \citep{Kutsenko23} the authors found that the active region can exhibit both acceleration and deceleration during occurrence, whereas during attenuation, the rotation rate remains virtually unchanged. There are variations in different cycles and phases within cycles, as shown by \citep{Balthasar1986}  and \citep{Howard}. The highest rotation rates are observed during the phases of minimum activity.

There are much fewer studies based on the analysis of the movement of individual sunspots. \cite{NN}, studying the movements  single unipolar recurrent sunspots  found the dependence of the sidereal velocity of rotations of the sine of latitude: $\omega(\theta)=A+B\cdot{\rm sin}^{2}\theta =14.38-2.38  {\rm sin}^{2}\theta$ deg/day. In \citep{Kutsenko21} the rotation rate was found for unipolar sunspots   $\omega(\theta)= 14.28-2.4  {\rm sin}^{2}\theta-2.24  {\rm sin}^{4}\theta$ deg/day. In \citep{Howard} it was found that sunspot groups   rotate  about 1\% slower than individual sunspots. In \citep{Permata} the average rotational velocity $\omega(\theta)=14.376+0.6 {\rm sin}^{2}\theta$ deg/day was found.  As the authors point out, the positive constant at ${\rm sin}^{2}\theta$ may be associated with errors in determining the rate.

\cite{Gilman84,Gilman85}  performed the analysis of the rotation rate variation of individual sunspots according to the Mount Wilson Observatory. They processed photographic plates of observations in “white” light for the period 1917\,--\,1983.  In \citep{Gilman85}  sunspots with a lifetime of at least 2 days were taken for analysis. It was found that the leading sunspots rotate faster than the following  sunspots, by about 0.1 deg/day, or 14 m/s.

It should be taken into account that the measured rotation rate may be superimposed by sunspots own movements. 	In this study, we will look at the rate of longitude movement of individual sunspots, focusing on different types of sunspots. Section 2 presents a description of the data, a method for identifying sunspots and determining the velocity of movement.  In Section 3.1, the velocity of longitude movement of sunspots of various types such as pores and regular sunspots is investigated. In section 3.2, the velocities for  sunspots of the leading and trailing polarity are considered. Section 4 presents a discussion of the results obtained.

\section{Data and Method}
For the analysis, we used  \textit{Solar Dynamics Observatory/Helioseismic and Magnetic Imager} (SDO/HMI), with a cadence of 45 seconds of the hmi.Ic\_45s and hdmi.M\_45s series made at the same time.  We took 5 images for each day at times close to 00:00, 05:00, 10:00, 15:00 and 20:00 UT (\url{http://jsoc.stanford.edu/ajax/exportdata.html}).  The main data for processing were images in “white” light (hmi.Ic\_45s), on which sunspots and sunspot umbra were highlighted.  The observation data were processed automatically. To identify sunspots and pores, we used a sunspot detection procedure similar to that used at the Kislovodsk Mountain Astronomical Station (KMAS) to identify sunspots for synoptic observations. The analysis of KMAS synoptic observations uses white light observations made with a 150 mm diameter photoheliograph. To extract sunspots, automatic sunspot segmentation procedures are used. But also an important feature in the processing program in Kislovodsk is the option where the operator can change methods, method parameters and perform object segmentation manually. This provides greater flexibility in the use of the program, which is especially important when processing observations with defects and in clouds \citep{Tlatov14, Tlatov19}.

Initially, the darkening towards the edge of the Sun was eliminated in the images of the Sun in the continuum, and the level of quiet Sun $I_{\rm{qs}}$ was determined.  
To eliminate limb darkening and determine the background intensity, the solar disk was divided into 12 segments by polar angle. In each segment, the dependence of intensity on radius was found. The intensity at each point depending on the polar angle and distance was found by interpolation \citep{Tlatov19}.
Then, regions of existence of sunspot groups of with a size of $\pm 6^{\rm{o}}$ in latitude and $\pm 8^{\rm{o}}$ degrees in longitude from the center of the sunspot groups were highlighted on the disk.  The coordinates of the sunspot groups were taken from the data \url{http://solarcyclescience.com/activeregions.html}.

\begin{figure}[ht!]
\centerline{\includegraphics[width=0.8\textwidth,clip=]{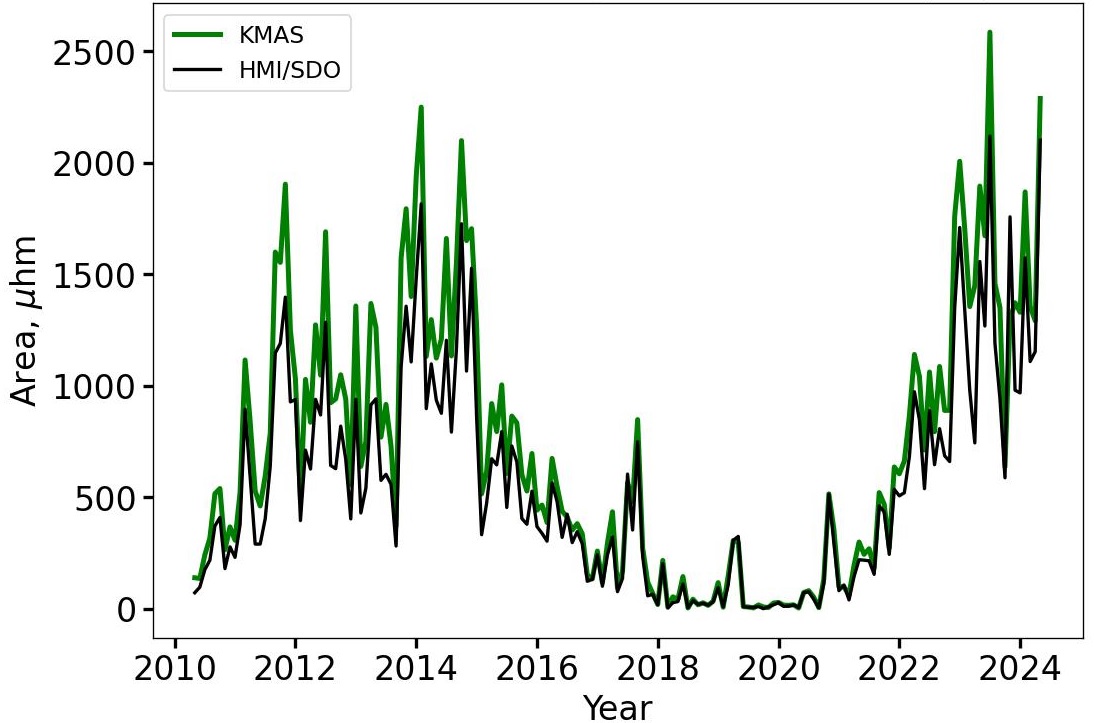}
                 } 
\caption{Comparison of monthly sunspot area data from KMAS synoptic ground observations and SDO/HMI automatic processing data.}
\label{fig:fig1}
\end{figure}

A sunspot search was conducted in these regions. To search for the photosphere-penumbra boundary $I_{\rm{pu}}$, the method of “growing” from a "seed" with a minimum intensity of  $I_{\rm{mn}}<0.92 I_{\rm{qs}}$ was used. The growing procedure was carried out in increments of $10^{\rm{-2}} I_{\rm{qs}}$. The area gain was calculated at each step. When reaching a level close to $I_{\rm{qs}}$, the increase became large.  The procedure was then stopped and the second to last step level was applied as the $I_{\rm{pu}}$ threshold.  Umbra and penumbra were allocated for each sunspot. The procedure for searching for the umbra-penumbra boundary $I_{\rm{up}}$ was similar, but instead of the $I_{\rm{qs}}$ intensity, the intensity of the penumbra-photosphere boundary $I_{\rm{pu}}$ was used.  The search started with points with an intensity lower than $0.92 I_{\rm{pu}}$.   If the penumbra could not be allocated, then such spots were marked as solar pores.

The technique of  sunspot detection is also described in \citep{Tlatov14, Tlatov22a}. To determine the polarity of the magnetic field, we superimposed the boundaries of  sunspots on the observations of magnetic fields (hdmi.M\_45s) and determined the maximum magnetic field along the line of sight $B_{\rm{LOS}}$. Pores in which the intensity of the magnetic field was $B_{\rm{LOS}}<100$ G were excluded from consideration \citep{Tlatov22a}.  In total, more than 87 thousand sunspots and more than 200 thousand pores were allocated in the period 01.05.2010-31.05.2024.  

The data of individual sunspots based on the results of processing ground-based observations in the period 2010\,--\,2024 are available in tabular form \url{//solarstation.ru/archive}   and in graphic form on the website \url{https://observethesun.com}.  Detailed differences between the characteristics of pores and regular sunspots are presented \citep{TlatovPevtsov, Tlatov19, Tlatov22a}.

Figure \ref{fig:fig1} shows the average monthly values of the area according to the KMAS ground-based synoptic observation processing data and the SDO/HMI automatic processing data used by us for analysis.  The correlation coefficient was r=0.98. Some differences in the data are mainly due to the difficulty of determining the area at the limb during the automatic procedure. Despite the significant number of pores in the total area, they account for approximately  15\%.

To determine the rotation rate, you need to track sunspots on time-adjacent images. To do this, we compared  sunspots using these characteristics, primarily coordinates, the sign of the polarity of the magnetic field and the area.  We searched for pairs of spots at various points in time at the minimum expected distance, taking into account the polarity of the magnetic field and the limitation on the difference in area.  In total, more than 210 thousand pairs were allocated in neighboring images of the Sun with a difference of 5 hours.  The sequence determination procedure applied to determining the lifetime of sunspots and pores is presented in \citep{Tlatov23}.
Next, we determined the rate of movement of the geometric center of sunspots and pores $\omega=\Delta \phi/\Delta t$, where $\Delta \phi$ is the longitude shift, $\Delta t$ is the time interval.

\begin{figure}
\centerline{\includegraphics[width=0.8\textwidth,clip=]{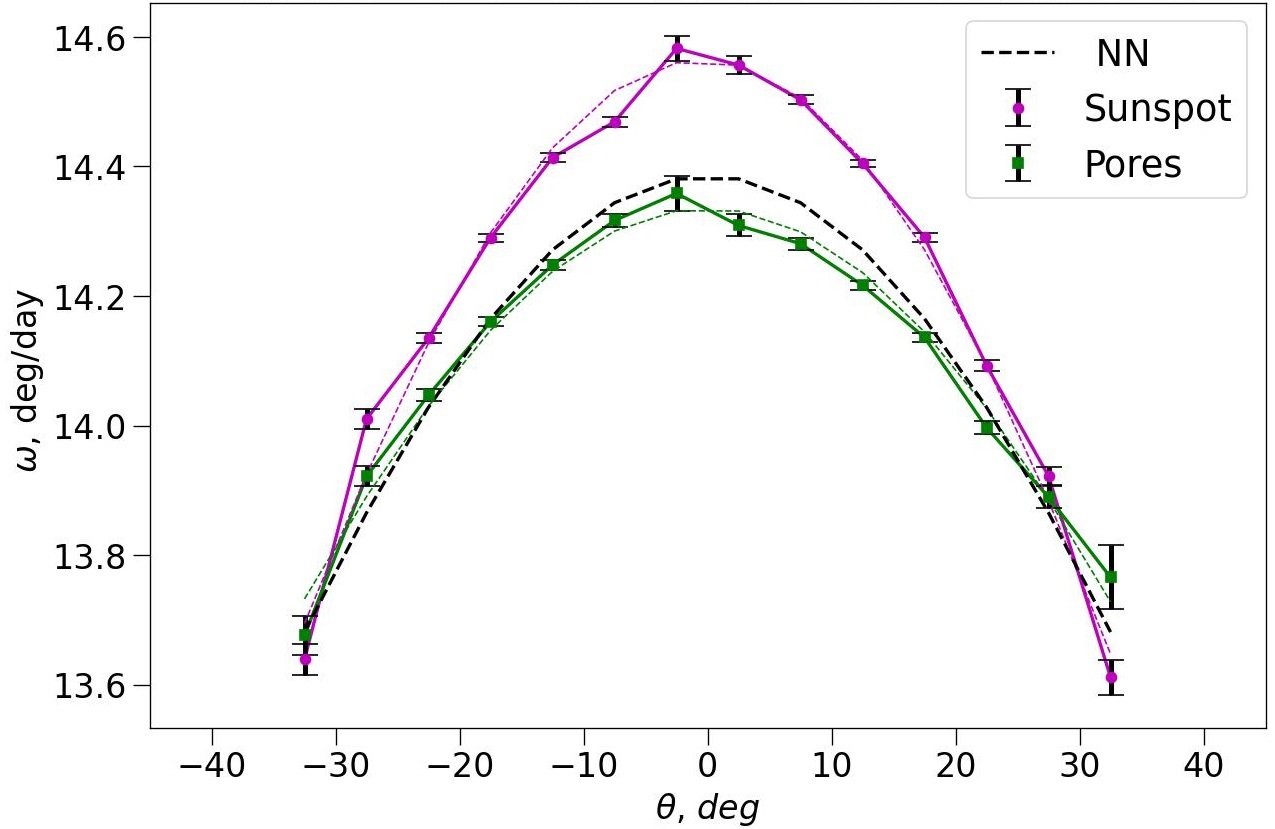}
                 } 
\caption{The rotation rate of individual sunspots and pores.  The black dotted line (NN) represents the rotation rate of  unipolar sunspots   \citep{NN}. A confidence interval for approximations is presented. The dotted lines represent approximations for different types of spots.}
\label{fig:fig2}
\end{figure}

\section{Analysis Results}
\subsection{Rotation of Sunspots and Solar Pores}

Sunspots can be divided into two types according to these morphology.  Regular sunspots that have umbra and penumbra, and pores (in which the penumbra cannot be distinguished). Pores have a much smaller area than regular sunspots. Pores form sunspot groups of class A and B according to the Zurich classification \citep{Tlatov19, Tlatova}. In \citep{TlatovPevtsov, Tlatov23} it was shown that the magnetic fields of pores and regular sunspots have different dependences on the area. This may indicate different physical processes in the formation of these types of spots.  Therefore, it is advisable to study the rotation of these two types of spots separately.

\begin{figure}
\centerline{\includegraphics[width=0.8\textwidth,clip=]{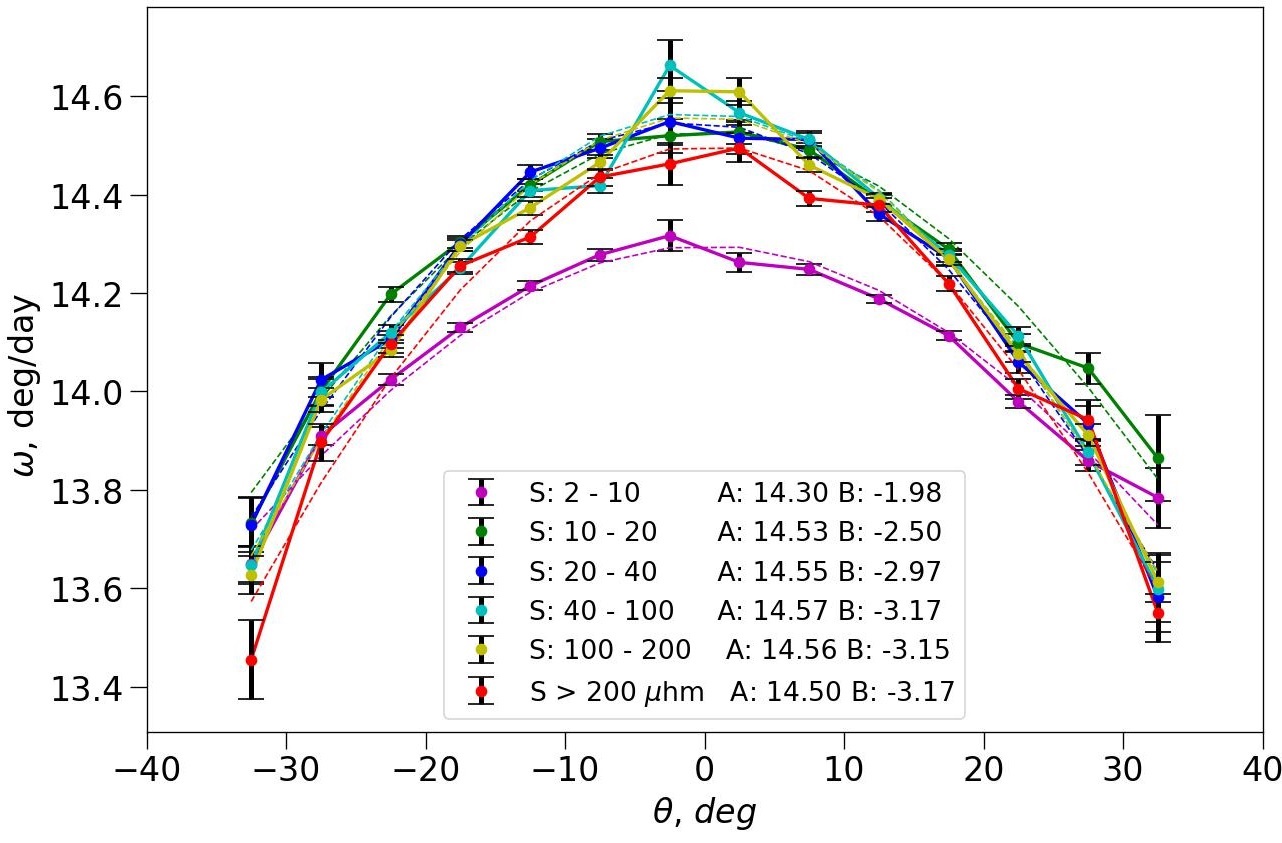} } 
\caption{Rotation rate of sunspots of different area intervals. The dotted lines represent approximations for spots of different areas.
The coefficients of approximations $\omega(\theta)=A+B\cdot{\rm sin}^{2}\theta $ are presented.
}
\label{fig:fig3}
\end{figure}

To study the dependence of rate on latitude, we looked for average rate values in latitude intervals of $5^{\rm o}$ degrees. Figure ~\ref{fig:fig2} shows the sidereal rotation rate of pores and sunspots.  Using the least squares method, we obtained the following dependencies for sunspots: 
$\omega_{sp}(\theta)= 14.56-3.094 \rm{sin}^{\rm 2}\theta-0.04 \rm{ sin}^{\rm 4}\theta$,  and for pores: $\omega_{pr}(\theta)= 14.34-2.097 \rm{sin}^{\rm 2}\theta-0.01 \rm{sin}^{\rm 4}\theta$ deg/day. 

Sunspots rotate faster than solar pores and have a greater degree of latitude differentiation.  The average pore area for which the rotation velocity was found was $S_{\rm pr} \approx 9.4$ $\mu$hm, and the average sunspot area $S_{\rm sp} \approx 133.6$ $\mu$hm. Thus, larger sunspot objects rotate faster than pores.

\subsection{Rotation of Sunspots of Different Area}
In this section, we will consider the change in the rotation rate from the area of sunspots and pores. There are contradictory results on the issue of changing the rotation rate from the area. 
To study the question of the rotation rate from the area, we divided all  sunspots into area intervals $S$ : 2\,--\,10, 10\,--\,20, 20\,--\,40, 40\,--\,100, 100\,--\,200, $>$200 $\mu$hm. For this analysis, we did not separate sunspots according to morphology, that is, into pores, sunspots without penumbra and regular sunspots. And only the area was taken into account.

Figure ~\ref{fig:fig3} shows graphs dependence of the rotation rate on latitude. The coefficients of approximation of coefficients $A$ and $B$ are also presented. It can be noted that, depending on the area, all the spots can be divided into two groups.  Small area spots $S<10$ $\mu$hm, which rotate relatively slowly and their rotation velocity at the equator is 14.30 deg/day. For larger area spots $S > 10$ $\mu$hm, the rotation velocity at the equator is $\approx 14.56$ deg/day and changes slightly with area.  The areas of the spots $S<10$ $\mu$hm usually correspond to the pores.
Figure \ref{fig:fig3} confirms the results of Figure \ref{fig:fig2}. 

\begin{figure}
\centerline{\includegraphics[width=0.8\textwidth,clip=]{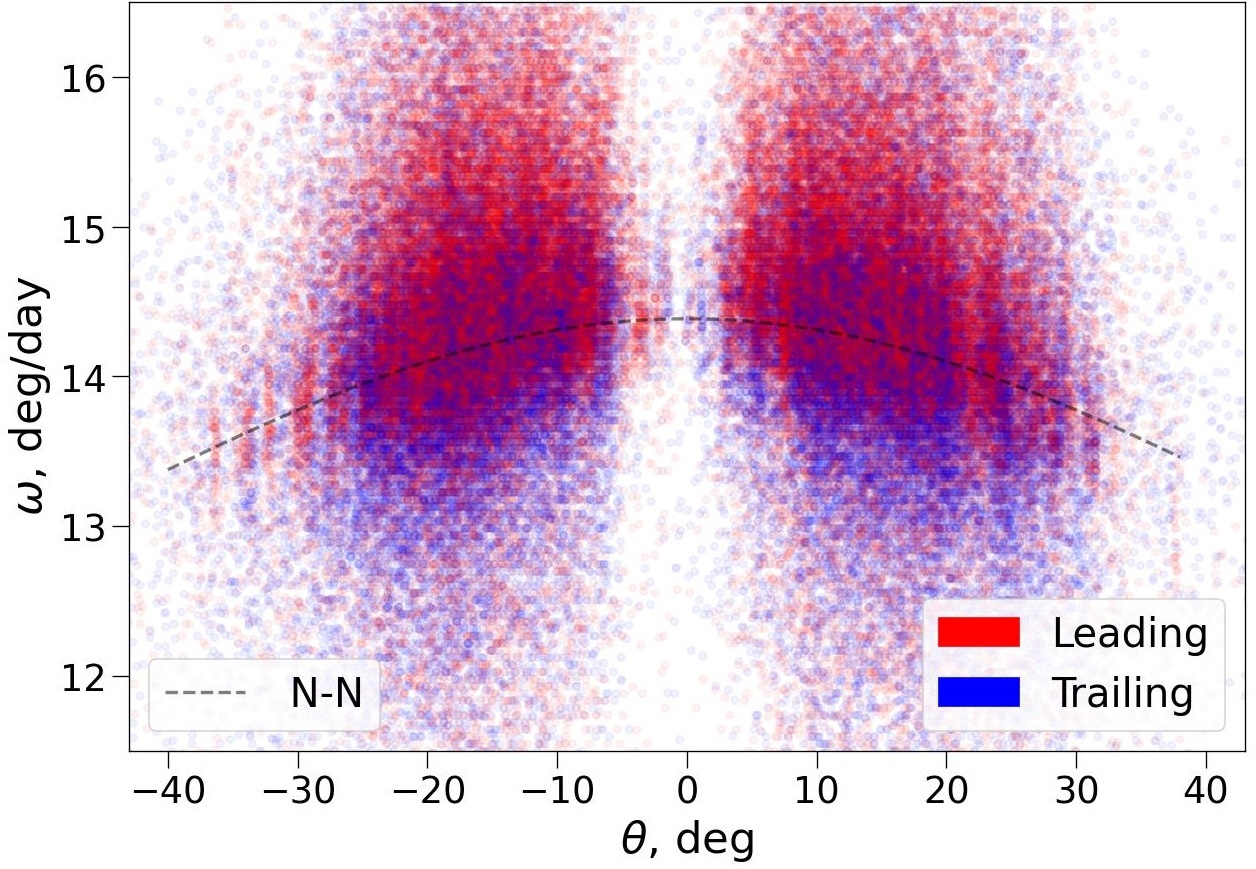} } 
\caption{The scattering diagram of  longitudinal velocity  of  leading and trailing polarity  sunspots.
}
\label{fig:fig4}
\end{figure}

\subsection{Rotation  of Sunspots of Leading and Trailing  Polarity}

Another sample of sunspots of various types may be the separation by the polarity of the magnetic field.  To determine the polarity of the field, we used the superposition of the boundaries of sunspots identified in the continuum on magnetograms taken at the same time. The Hale's law was used to determine the polarity sign of the leading and trailing sunspots.

In total, $\approx 117$ thousand spots of the leading polarity and $\approx 95$ thousand of the trailing polarity were identified, in which the velocity of longitude movement was determined.  Figure \ref{fig:fig4} shows a scattering diagram of the velocity of longitude movement of regular sunspots of different polarities.  Leading polarity sunspots are shown in red, trailing polarity sunspots are shown in blue.  We can note that the graph shows the separation of the velocity of movement for these types of spots

\begin{figure}
\centerline{\includegraphics[width=0.8\textwidth,clip=]{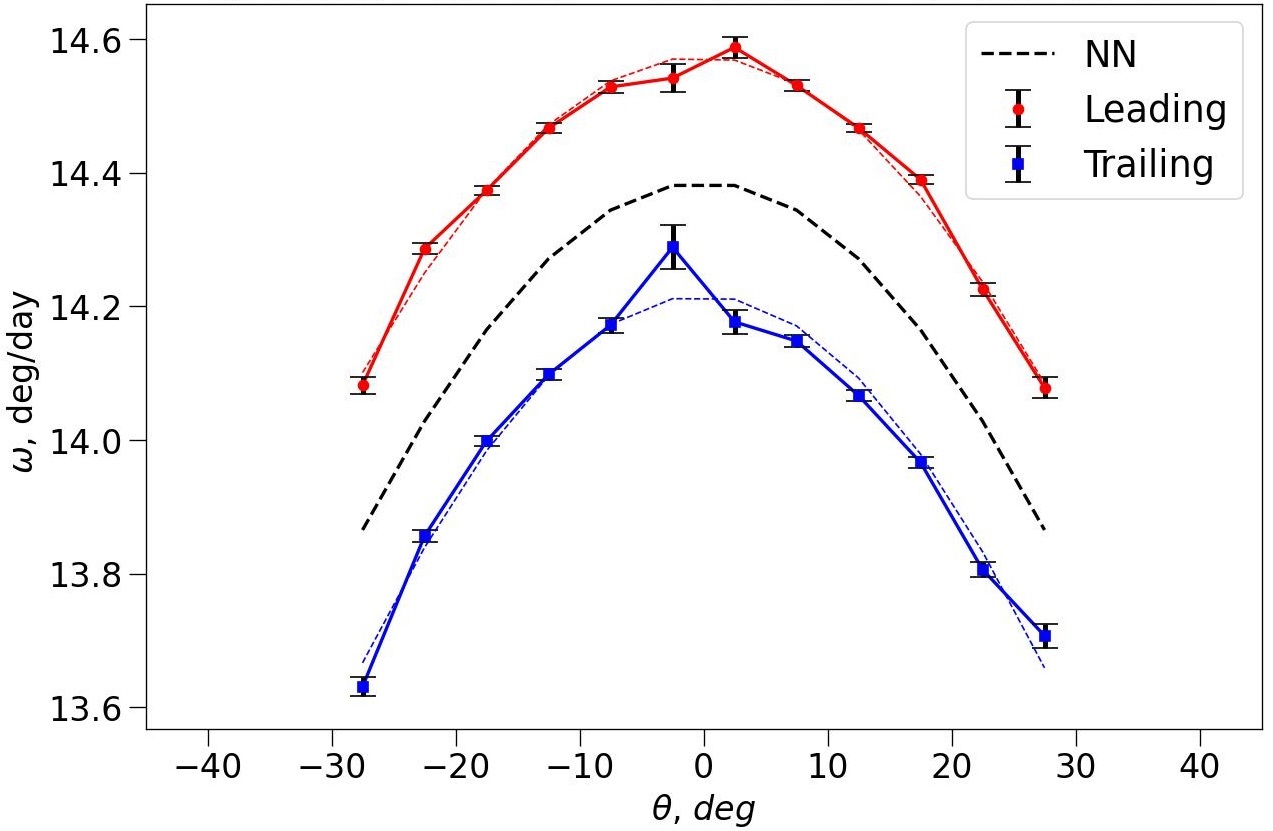} } 
\caption{The rotation rate of  sunspots and pores separately of  leading and trailing polarities.
The dotted lines represent approximations for spots of different polarity.
}
\label{fig:fig5}
\end{figure}

Figure \ref{fig:fig5} shows approximations of the displacement velocity for sunspots and pores of the leading and trailing polarities. For sunspots of the leading polarity, the rotation rate can be approximated by the formula:
 $\omega_{\rm ld}(\theta)=14.574-2.225 \rm{sin}^{\rm 2}\theta-0.02 \rm{ sin}^{\rm 4}\theta$  deg/day. For trailing polarity spots: $\omega_{\rm tr}(\theta)=14.216-2.60 \rm{sin}^{\rm 2}\theta-0.01 \rm{ sin}^{\rm 4}\theta$ deg/day.  Sunspots of the leading polarity rotate $\approx 2.4\%$ faster than trailing polarity sunspots.

\begin{figure}
               \includegraphics[width=0.48\textwidth,clip=]{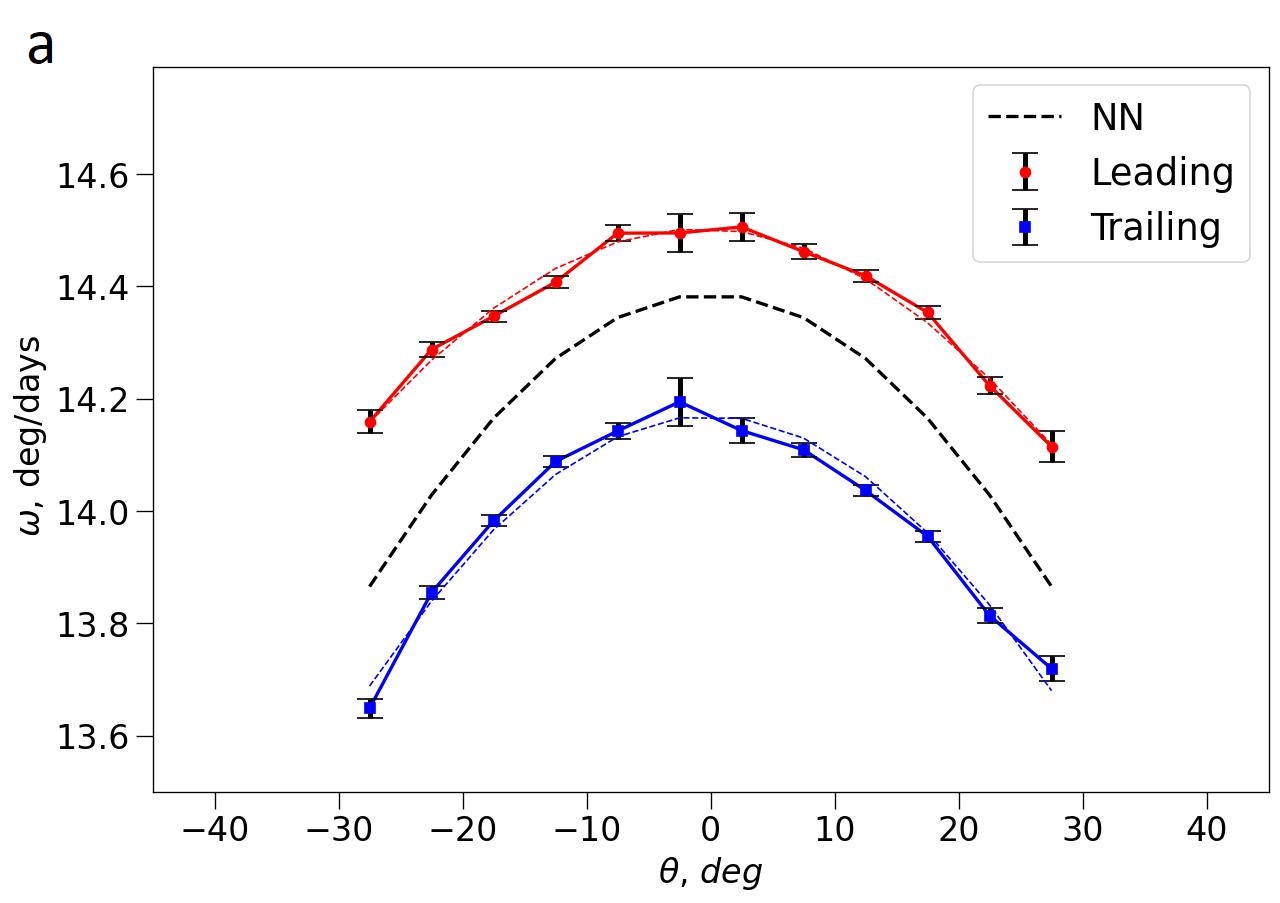}
               \hspace*{0.013\textwidth}
               \includegraphics[width=0.48\textwidth,clip=]{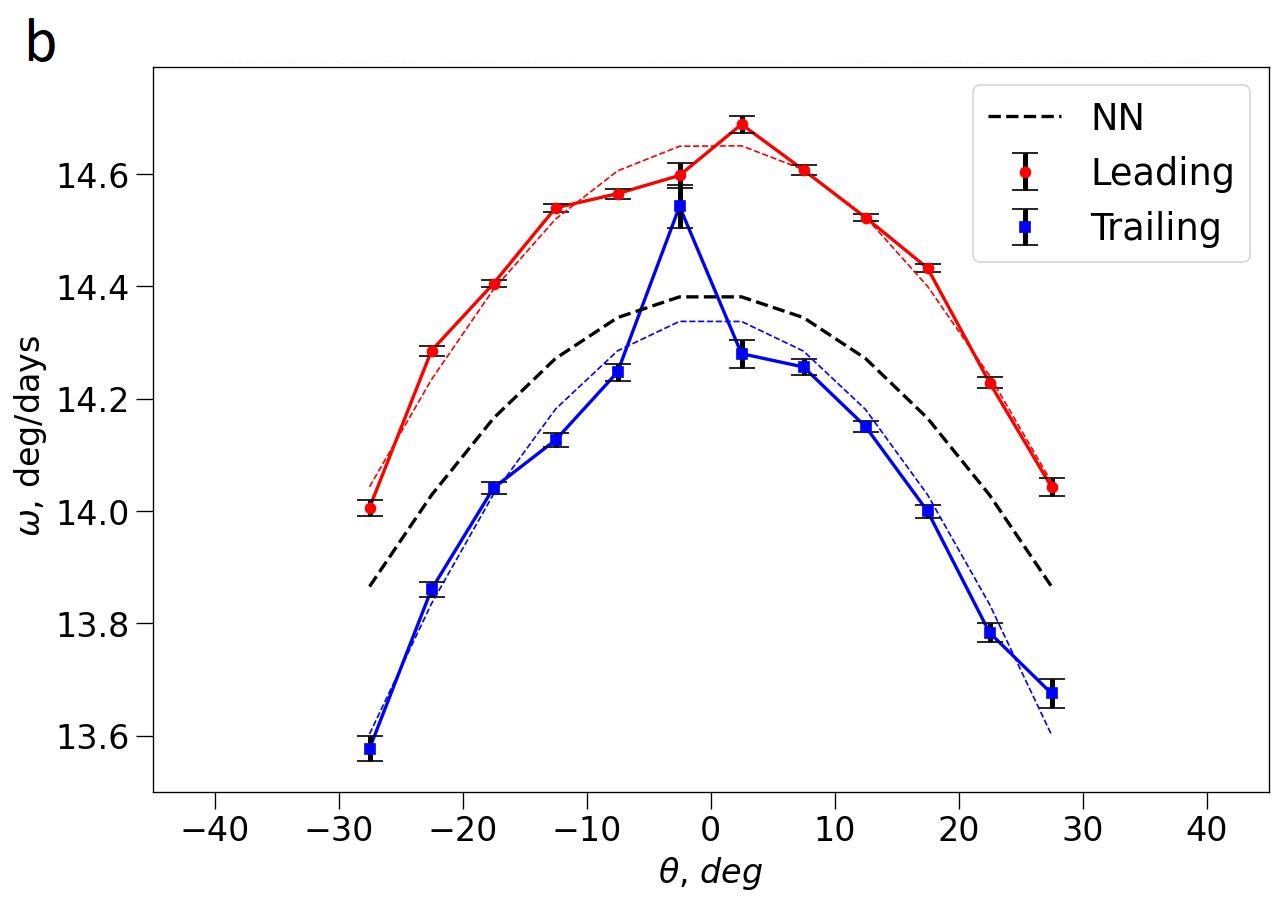}
\caption{The same as Figure \ref{fig:fig5}, but separately for pores (left) and sunspots (right).   
}
\label{fig:fig6}
\end{figure}

Figures \ref{fig:fig6}   shows the dependences of the rotation rate of the leading and trailing polarity spots separately for different types of spots: pores and sunspots. For pores, this dependence can be represented as:  $\omega^{ld}_{pr}(\theta)= 14.50-1.71 \rm{sin}^{\rm 2}\theta $ deg/day  for leading polarity; and  $\omega^{tr}_{pr}(\theta)= 14.17-2.28 \rm{sin}^{\rm 2}\theta $ deg/day for trailing polarity. 
For sunspots, the dependencies are as follows: $\omega^{ld}_{sp}(\theta)= 14.65-2.85 \rm{sin}^{\rm 2}\theta $ deg/day leading polarity; and $\omega^{tr}_{sp}(\theta)= 14.34-3.48 \rm{sin}^{\rm 2}\theta$ deg/day  for trailing polarity.

Again, we can note that the pore rotation rate is lower than the rotation rate of sunspots, both for spots of the leading and trailing polarity.

Near the equator in Figures \ref{fig:fig5}, \ref{fig:fig6}  there are deviations from the patterns. This is due to errors in determining the leading polarity in the hemispheres.  It is possible that due to random processes, sunspots and pores cross the equator and thereby violate Halle's law of polarity for the hemispheres.  We took such sunspots to be spots of leading or trailing polarity. But at the same time, these rotation rate corresponded to sunspots from the opposite hemisphere.

\begin{figure}
\centerline{\includegraphics[width=0.8\textwidth,clip=]{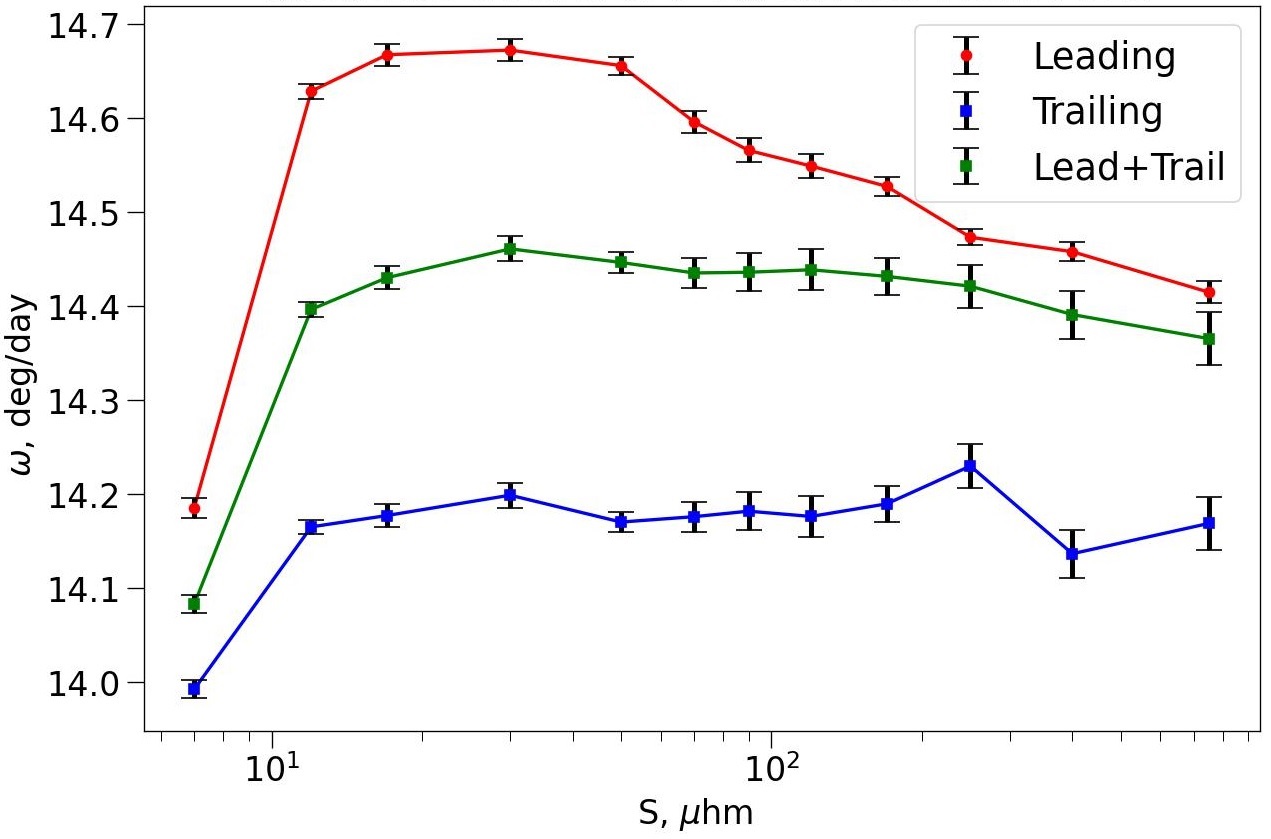} } 
\caption{Rotation rate of spots of different magnetic polarity depending on the area for spots with latitude in the range $\theta<\pm 15^o$.   
}
\label{fig:fig7}
\end{figure}

\begin{table}
\centering
\tabletypesize{\scriptsize}
\tablewidth{700pt} 

\caption{Summary of sunspots and pores sidereal rotation rates (deg/day).
}
\label{T-simple}

\begin{tabular}{lc} 

\hline
 \multicolumn{2}{c}{Pores} \\
 \hline

all & $\omega_{pr}(\theta)= 14.34-2.01 \rm{sin}^{\rm 2}\theta $ \\
Leading  & $ \omega^{ld}_{pr}(\theta)= 14.50-1.71 \rm{sin}^{\rm 2}\theta$ \\
Trailing & $\omega^{tr}_{pr}(\theta)= 14.17-2.286 \rm{sin}^{\rm 2}\theta$ \\
\hline
 \multicolumn{2}{c}{Sunspots } \\
 \hline
all & $\omega_{sp}(\theta)= 14.56-3.01 \rm{sin}^{\rm 2}\theta$ \\
Leading  & $ \omega^{ld}_{sp}(\theta)= 14.65-2.85 \rm{sin}^{\rm 2}\theta$ \\
Trailing & $\omega^{tr}_{sp}(\theta)= 14.34-3.48 \rm{sin}^{\rm 2}\theta$ \\

\hline
 \multicolumn{2}{c}{Area} \\
 \hline
$S <10$ $\mu$hm   & $\omega(\theta)= 14.30-1.98 \rm{sin}^{\rm 2}\theta $ \\
$ 10\,-\,20$ $\mu$hm   & $\omega(\theta)= 14.53-2.50 \rm{sin}^{\rm 2}\theta$ \\
$ 20\,-\, 40$ $\mu$hm   & $\omega(\theta)= 14.55-2.97 \rm{sin}^{\rm 2}\theta$ \\
$ 40\,-\,100$ $\mu$hm  & $\omega(\theta)= 14.57-3.17 \rm{sin}^{\rm 2}\theta$ \\
$ 100\,-\,200$ $\mu$hm  & $\omega(\theta)= 14.56-3.15 \rm{sin}^{\rm 2}\theta$ \\
$S >200$ $\mu$hm  & $\omega(\theta)= 14.51-3.17 \rm{sin}^{\rm 2}\theta$ \\

\hline
\end{tabular}
\end{table}

The dependence of the differential rotation rate of sunspots of different magnetic polarity on the area is shown in Figure \ref{fig:fig7}. Here, spots near the equator $\theta<\pm 15^o$ are considered. The following can be noted. Small sunspots with an area of up to $S<10$ $\mu$hm rotate more slowly than spots with a larger area. For sunspots with an area of S:10-50 µhm, the rotation rate practically does not change with area. For sunspots of leading polarity, with increasing area at $S>50$ $\mu$hm, the rotation rate decreases. But for spots of trailing polarity, the dependence of the rate on the area is not visible.

\section{Discussion}

We performed the analysis of the longitude movement of individual sunspots. We found that the velocity of movement of sunspots significantly depends on these type.  Table 1 shows the values of approximations of rotation rates from the area and for various types of sunspots. 
The separation of sunspots into pores, that is, spots without penumbra, and regular sunspots with penumbra, showed that regular sunspots rotate $\approx1.5\%$ faster than pores  ($A_{\rm pr}=14.34$, $A_{\rm sp}=14.56$ deg/day).  At the same time, the degree of differential rotation of sunspots is $\approx 47\%$ higher than for pores ($B_{\rm pr}=2.1$, $B_{\rm sp}=3.09$ deg/day) (Figure ~\ref{fig:fig2}). This result is confirmed by the analysis of the rotation rate of sunspots of various areas (Figure ~\ref{fig:fig3}). Small spots $S < 10$ $\mu$hm rotate slowly. 

The rotation rateof sunspots with an area of $S > 10$ $\mu$hm is practically independent of the area (Figure ~\ref{fig:fig3}). This contradicts the conclusions of previous studies \citep{Howard, Gupta, Nagovitsyn} for the rotation of sunspot  groups. It is possible that there are errors in determining the rotation rate of sunspot groups. Consider measuring the rate of sunspot  groups  near the limbs, where measurements are difficult due to the effects of projection and the Wilson effect \citep{Wilson} . On the eastern limb, the average coordinate of the groups is shifted towards the leading sunspot during measurements. On the western limb, the average measured longitude of the sunspot group shifts to the trailing sunspots.   This means that the longitude distance increases during measurements, which leads to errors in measuring rotation rate. The greater the longitude of the group, the greater the error may be. Since the longitude and size are interrelated, this leads to an error in determining the rotation rate, which increases with the area of the group. Another factor influencing velocity measurement errors may be the faster decay of trailing regions \citep{Petrovay}.

\begin{figure}
\centerline{\includegraphics[width=0.8\textwidth,clip=]{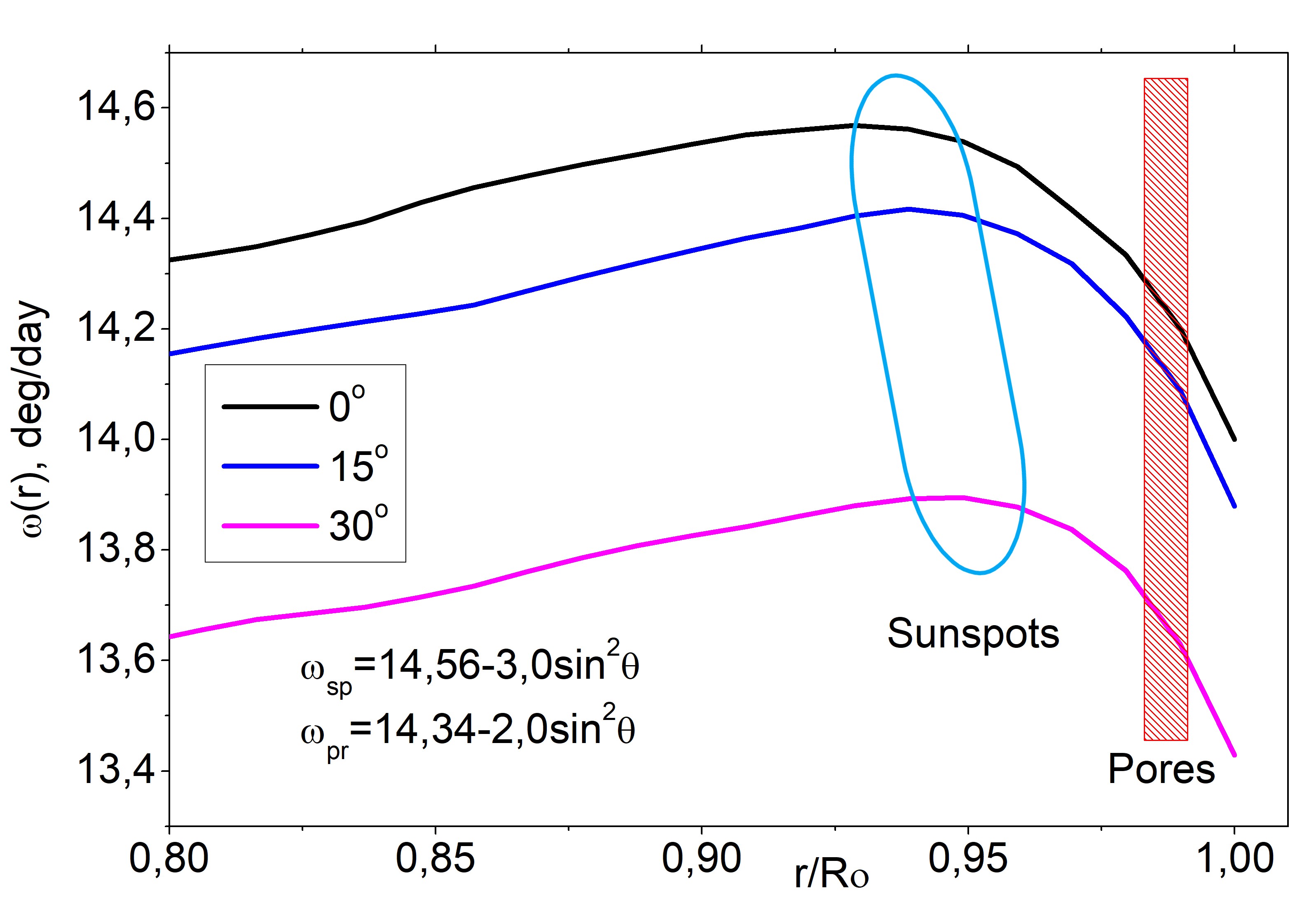} } 
\caption{Scheme of formation of sunspots and pores at different depths for latitudes $0^o$, $15^o$, $30^o$ inside the leptocline ($r \gtrsim 0.95R_{\odot}$). Rotation data are taken from \citep{Howe}.
}
\label{fig:fig8}
\end{figure}

The difference in the rotation rate of different types of sunspots is often explained within the framework of the anchor hypothesis. This hypothesis assumes that the magnetic flux beam is fixed at some depth in the convection zone \citep{Gilman84,Rhodes,Kutsenko23}. As a result, the rotation rate on the surface is determined by the rotation rate of the plasma at the depth of fixation. It is known that in the upper layer of the solar convective zone, called the leptocline ($r>0.95R_{\odot}$), the rotation rate changes rapidly with depth \citep{Howe}. At depths of $r>0.95R_{\odot}$, the rotation rate is maximum.  Pores have a much smaller area than regular sunspots. In our study, the average pore area for which the rotational rate was found in neighboring images was $S_{\rm pr}\approx 10$ $\mu$hm, or diameter $d_{\rm pr}\approx 1.954S^{\rm 0.5} \approx 6$ Mm, and the average sunspot area $S_{\rm sp}\approx 133$ $\mu$hm, or $d_{\rm sp}\approx 20$ Mm.  A comparison of the rotational velocities of different types of spots found in this study (Figure \ref{fig:fig2}) with the results of the rotation rate in depth \citep{Howe} shows that the rotation  of the pores is close to the region of delayed rotation in the leptocline at a depth of  $h_{\rm pr} \approx 0.98R_{\odot} \approx 14$ Mm. The rotation rate of regular sunspots is close to the values of the maximum rotation  at a depth of $h_{\rm sp} \approx 0.95R_{\odot} \approx 35$ Mm (Figure ~\ref{fig:fig8}).  It is possible that sunspots and pores are predominantly formed with a ratio of diameter to depth of the spot $h/d \approx 2$. Since the region of maximum rotation varies with depth (Figure ~\ref{fig:fig8}), it is possible that the degree of differential rotation for regular sunspots is therefore higher than for pores.

\begin{figure}
\centerline{\includegraphics[width=0.8\textwidth,clip=]{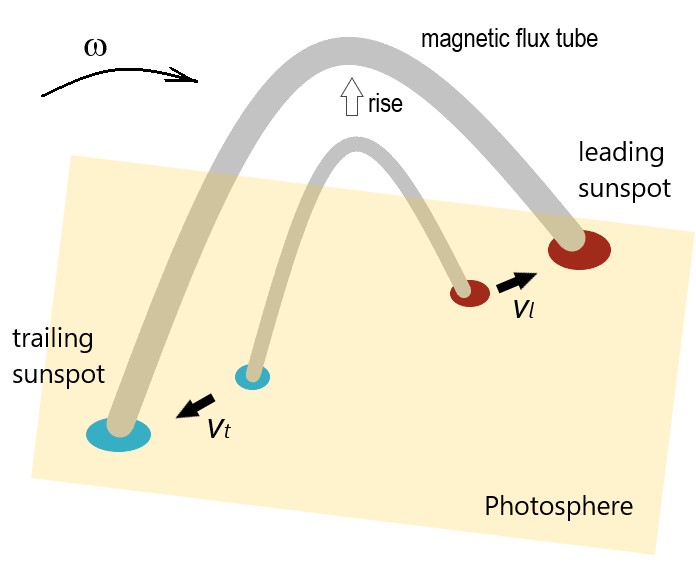} } 
\caption{A simplified scheme of the possible formation of an additional velocity of the longitudinal displacement of sunspots during the ascent of a magnetic flux tube.}
\label{fig:fig9}
\end{figure}

The second result of our study indicates that the velocity of the longitudinal movement of  sunspots depends on the polarity of the magnetic field. To separate the spots, we used the polarity of the magnetic field. A similar result was obtained earlier \citep{Gilman85}. To separate sunspots, the authors employed the leading  the leading and trailing sunspots of a sunspot groups  when processing the plates in white-light.   Sunspots and pores of the leading polarity have rotation rate of $\approx 2.4\%$, or 0.26 deg/day, or 36 m/s faster than the trailing polarity spots (Figures \ref{fig:fig5}). Such a difference in velocities may be due to additional factors affecting the horizontal movement of the spots.  Sunspots are visible as the exits of the arch of magnetic flux tubes on the photosphere. At the initial stage of ascent, the distance between sunspots is lower than in the later stages, since the distance between the legs of the arches fixed on the photosphere increases with distance from its top. This leads to the appearance of an additional observable velocity $v_{\rm l}$ for the leading spot directed along the rotation. For the trailing spots of the part, the $v_{\rm t}$ velocity is directed against rotation, which creates the effect of a slow rotation rate (Figure \ref{fig:fig9}).  It is likely that the rate of ascent and therefore the rate of longitudinal  flow differs for pores and regular sunspots. In Figure \ref{fig:fig6}, we see that the velocity differences are greater for pores than for regular sunspots. This corresponds to the hypothesis of the ascent of the flux tube. At the initial stage of the ascent, the apparent velocity of the run-up of the bases of the arches of the magnetic  flux tube will be greater, as can be seen from Figure \ref{fig:fig6}.

The hypothesis of the influence of the ascent of the magnetic flux tube on the velocity of longitudinal flows corresponds to anomalies in the rotation rate near the equator (Figure ~\ref{fig:fig6}). If, due to turbulence, the trailing or leading part of the bipolar group, one hemisphere, enters the other hemisphere, then we will mistakenly consider them leading or trailing according to Hale's law. But these rate will differ from the pattern for this hemisphere.

Studies of the longitudinal movement of individual sunspots and pores can help in determining the distribution of rotation rate with depth and can be used in models of dynamos and the formation of magnetic field flux tubes.

\begin{acknowledgments}
We acknowledge the financial support the Ministry of Science and Higher Education of the Russian Federation, grant number 075-03-2024-113.

\end{acknowledgments}

\bibliography{spot_rot.bib}{}

\begin{thebibliography}{}
\expandafter\ifx\csname natexlab\endcsname\relax\def\natexlab#1{#1}\fi
\providecommand{\url}[1]{\href{#1}{#1}}
\providecommand{\dodoi}[1]{doi:~\href{http://doi.org/#1}{\nolinkurl{#1}}}
\providecommand{\doeprint}[1]{\href{http://ascl.net/#1}{\nolinkurl{http://ascl.net/#1}}}
\providecommand{\doarXiv}[1]{\href{https://arxiv.org/abs/#1}{\nolinkurl{https://arxiv.org/abs/#1}}}

\bibitem[{{Balthasar} {et~al.}(1982){Balthasar}, {Schuessler}, \& {Woehl}}]{Balthasar}
{Balthasar}, H., {Schuessler}, M., \& {Woehl}, H. 1982, \solphys, 76, 21, \dodoi{10.1007/BF00214127}

\bibitem[{{Balthasar} {et~al.}(1986){Balthasar}, {Vazquez}, \& {Woehl}}]{Balthasar1986}
{Balthasar}, H., {Vazquez}, M., \& {Woehl}, H. 1986, \aap, 155, 87

\bibitem[{{Beck}(2000)}]{Beck}
{Beck}, J.~G. 2000, \aaps, 191, 47, \dodoi{10.1023/A:1005226402796}

\bibitem[{{Gilman} \& {Howard}(1984)}]{Gilman84}
{Gilman}, P., \& {Howard}, R. 1984, \aj, 283, 385, \dodoi{10.1086/162316}

\bibitem[{{Gilman} \& {Howard}(1985)}]{Gilman85}
---. 1985, \aj, 295, 233, \dodoi{10.1086/163368}

\bibitem[{{Gupta} {et~al.}(1999){Gupta}, {Sivaraman}, \& {Howard}}]{Gupta}
{Gupta}, S., {Sivaraman}, K.~R., \& {Howard}, F. 1999, \solphys, 295, 225, \dodoi{10.1023/A:1005229124554}

\bibitem[{{Howard} {et~al.}(1984){Howard}, {Gilman}, \& {Gilman}}]{Howard}
{Howard}, R., {Gilman}, P., \& {Gilman}, P. 1984, \aj, 283, 373, \dodoi{10.1086/162315}

\bibitem[{{Howe} {et~al.}(2000){Howe}, {Christensen-Dalsgaard}, {Hill}, {Komm}, {Larsen}, {Schou}, {Thompson}, \& {Toomre}}]{Howe}
{Howe}, R., {Christensen-Dalsgaard}, J., {Hill}, F., {et~al.} 2000, Science, 287, 2456, \dodoi{10.1126/science.287.5462.2456}

\bibitem[{{Javaraiah} \& {Gokhale}(1997)}]{Javaraiah}
{Javaraiah}, J., \& {Gokhale}, M.~H. 1997, \aap, 327, 795

\bibitem[{{Kutsenko}(2021)}]{Kutsenko21}
{Kutsenko}, A.~S. 2021, \mnras, 500, 5159, \dodoi{10.1093/mnras/staa3616}

\bibitem[{{Kutsenko} {et~al.}(2023){Kutsenko}, {Abramenko}, \& {Litvishko}}]{Kutsenko23}
{Kutsenko}, A.~S., {Abramenko}, V., \& {Litvishko}, D. 2023, \mnras, 519, 5315, \dodoi{10.1093/mnras/stac3826}

\bibitem[{{Nagovitsyn} {et~al.}(2018){Nagovitsyn}, {Pevtsov}, \& {Osipova}}]{Nagovitsyn}
{Nagovitsyn}, Y.~A., {Pevtsov}, A.~A., \& {Osipova}, A.~A. 2018, Astronomy Letters, 44, 202, \dodoi{10.1134/S1063773718020056}

\bibitem[{{Newton} \& {Nunn}(1951)}]{NN}
{Newton}, H.~W., \& {Nunn}, M. 1951, \mnras, 111, 413, \dodoi{10.1093/mnras/111.4.413}

\bibitem[{{Permata} \& {Herdiwijaya}(2019)}]{Permata}
{Permata}, K., \& {Herdiwijaya}, D. 2019, Journal of Physics: Conference Series, 1231, article id. 012019, \dodoi{10.1088/1742-6596/1231/1/012019}

\bibitem[{{Petrovay}(1993)}]{Petrovay}
{Petrovay}, K. 1993, The magnetic and velocity fields of solar active regions. Astronomical Society of the Pacific Conference Series

\bibitem[{{Pulkkinen} \& {Tuominen}(1998)}]{Pulkkinen}
{Pulkkinen}, P., \& {Tuominen}, I. 1998, \aap, 332, 748

\bibitem[{{Rhodes} {et~al.}(1990){Rhodes}, {Cacciani}, {Korzennik}, {Tomczyk}, {Ulrich}, \& F.}]{Rhodes}
{Rhodes}, E.~J., {Cacciani}, A., {Korzennik}, S., {et~al.} 1990, \apj, 351, 687, \dodoi{10.1086/168507}

\bibitem[{{Sivaraman} {et~al.}(2003){Sivaraman}, Sivaraman, {Gupta}, \& F.}]{Sivaraman}
{Sivaraman}, K.~R., Sivaraman, H., {Gupta}, S.~S., \& F., H.~R. 2003, \solphys, 214, 65, \dodoi{10.1023/A:102407510066}

\bibitem[{{Tlatov}(2022a)}]{Tlatov22a}
{Tlatov}, A. 2022a, \solphys, 297, article id.67, \dodoi{10.1007/s11207-022-02002-8}

\bibitem[{{Tlatov}(2023)}]{Tlatov23}
---. 2023, \solphys, 298, article id.93, \dodoi{110.1007/s11207-022-02045-x}

\bibitem[{{Tlatov} \& {Pevtsov}(2014)}]{TlatovPevtsov}
{Tlatov}, A., \& {Pevtsov}, A. 2014, \solphys, 289, 1143, \dodoi{10.1007/s11207-013-0382-9}

\bibitem[{{Tlatov} {et~al.}(2014{\natexlab{a}}){Tlatov}, {Riehokainen}, \& {Tlatova}}]{Tlatov19}
{Tlatov}, A., {Riehokainen}, A., \& {Tlatova}, K. 2014{\natexlab{a}}, \solphys, 294, article id. 45, \dodoi{10.1007/s11207-019-1439-1}

\bibitem[{{Tlatov} {et~al.}(2014{\natexlab{b}}){Tlatov}, {Vasil’eva}, {Makarova}, \& {Otkidychev}}]{Tlatov14}
{Tlatov}, A., {Vasil’eva}, V., {Makarova}, V., \& {Otkidychev}, P. 2014{\natexlab{b}}, \solphys, 289, 1403, \dodoi{10.1007/s11207-013-0404-7}

\bibitem[{{Tlatova} {et~al.}(2022){Tlatova}, {Vasil'eva}, {Berezin}, {Illarionov}, \& {Tlatov}}]{Tlatova}
{Tlatova}, K., {Vasil'eva}, V., {Berezin}, I., {Illarionov}, E., \& {Tlatov}, A. 2022, Astronomy Reports, 66, 165, \dodoi{10.1134/S1063772922030052}

\bibitem[{{Tuominen}(1962)}]{Tuominen}
{Tuominen}, J. 1962, Zeitschrift für Astrophysik, 55, 110

\bibitem[{{Wilson} \& {Maskelyne}(1774)}]{Wilson}
{Wilson}, A., \& {Maskelyne}, N. 1774, Philosophical Transactions (1683-1775), 64, 1

\end{thebibliography}
\bibliographystyle{aasjournal}



\end{document}